\documentstyle[prd,aps]{revtex}

\newcommand{\bea}{\begin{eqnarray}}
\newcommand{\eea}{\end{eqnarray}}

\begin{document}

\draft
\twocolumn[\hsize\textwidth\columnwidth\hsize\csname
@twocolumnfalse\endcsname

\title{Cosmological Gravitational Wave in a Gravity\\
       with Quadratic Order Curvature Couplings}
\author{Hyerim Noh}
\address{Korea Astronomy Observatory,
         San 36-1, Whaam-dong, Yusung-gu, Daejon, Korea}
\author{Jai-chan Hwang}
\address{Department of Astronomy and Atmospheric Sciences,
         Kyungpook National University, Taegu, Korea}
\date{\today}
\maketitle

\begin{abstract}

We present a set of equations describing the cosmological gravitational wave 
in a gravity theory with quadratic order gravitational coupling terms 
which naturally arise in quantum correction procedures.
It is known that the gravitational wave equation in the gravity theories 
with a general $f(R)$ term in the action leads to a second order 
differential equation with the only correction factor appearing in 
the damping term.
The case for a $R^{ab} R_{ab}$ term is completely different. 
The gravitational wave is described by a fourth order differential 
equation both in time and space.
However, curiously, we find that the contributions to the background
evolution are qualitatively the same for both terms.

\end{abstract}

\vskip2pc]

{\it 1. Introduction:}
Quantum correction procedures generate higher order curvature coupling 
terms in the action for Einstein gravity \cite{QFCS}. 
Such corrections are needed for more realistic understanding of the
early stage of the universe where the curvature of the universe was 
large enough for the quantum effects.
It is known that to one-loop order Einstein action is modified by 
the additional quadratic order curvature coupling terms \cite{QFCS}.
In four dimensional spacetime, we have two such correction terms: 
$R^2$ and $R^{ab}R_{ab}$ terms.
The first term has attracted much attention in the literature in the context
of cosmology and the cosmological structure formation theory based on 
linear gravitational stability analyses.
Inflationary stage was first implemented in $R^2$ gravity
before it was realized based on the scalar field \cite{Starobinsky}.
Now, it is well known that through a conformal transformation of 
the metric $R^2$ gravity can be transformed to a minimally coupled 
scalar field \cite{CT}.
Thus, despite its quadratic nature, it has been shown that the dynamical 
equations describing the structures in the cosmological model remain to 
the second order differential equation both for the scalar type 
perturbations and for the gravitational wave 
\cite{GGT-1990,H-PRW,GGT-others,GGT-CT}.
Such a symmetry does not exist in the $R^{ab}R_{ab}$ term.

In a series of papers we will investigate the roles of the $R^{ab} R_{ab}$ 
term in the context of the evolution of the cosmological structures.
In the present work we will derive the equations for the background 
and the gravitational wave.
The main results are Eqs. (\ref{AB-eq-BG},\ref{AB-eq-GW}).
Since the cosmological stability analysis for $R^{ab} R_{ab}$ term is new, 
we present the necessary details for the derivation in the Appendices.
We take MTW \cite{MTW} convention except for setting
$c \equiv 1 \equiv 8 \pi G$.

{\it 2. Gravity theory with the quadratic order quantum correction terms:}
We considered a gravity theory represented by the following action 
\bea
   S = \int d^4 x \sqrt{-g} \left[ {1 \over 2} \left(
       R + A R^2 + B R^{ab} R_{ab} \right) + L_m \right],
   \label{action}
\eea
where $L_m$ is the matter part Lagrangian.
The gravitational field equation becomes
\bea
   & & R^a_b - {1 \over 2} \delta^a_b R
       - A H^{(1)a}_{\;\;\;\;\;b} - B H^{(2)a}_{\;\;\;\;\;b} = T^a_b,
   \nonumber \\
   & & 
       H^{(1)a}_{\;\;\;\;\;b} = 2 R^{;a}_{\;\;\;b} - 2 \delta^a_b \Box R
       + {1 \over 2} \delta^a_b R^2 - 2 R R^a_b,
   \label{GFE} \\
   & &
       H^{(2)a}_{\;\;\;\;\;b}
       = R^{;a}_{\;\;\; b} - \Box R^a_b - {1 \over 2} \delta^a_b \Box R
       + {1 \over 2} \delta^a_b R^c_d R^d_c + 2 R^{cd} R^a_{\;\;cdb}.
   \nonumber
\eea
In four dimensional spacetime, due to the generalized Gauss-Bonnet theorem,
contributions from $R^{abcd} R_{abcd}$ term can be expressed in terms of
$H^{(1)a}_{\;\;\;\;\;b}$ and $H^{(2)a}_{\;\;\;\;\;b}$ terms.
These modifications of Einstein gravity were first introduced
by Weyl, Pauli and Eddington \cite{Weyl}.
Not only these terms arise from the one-loop level quantum correction,
these are also known to make the theory renormalizable \cite{Stelle}.

{\it 3. The cosmological model and the gravitational wave:}
As the model describing the background universe we consider a homogeneous, 
isotropic and {\it flat} model.
The general perturbation in this background model can be decomposed
into three types: the scalar, vector and tensor type perturbations.
To the linear order, due to the symmetry in the background, these three types 
of perturbations decouple from each other and evolve independently.
In this paper we will consider only the tensor type perturbation which is 
transverse and tracefree; the other types of perturbations will be 
considered elsewhere.
The metric can be written as
\bea
   ds^2 = - a^2 d \eta^2 + a^2 \left( \delta_{\alpha\beta}
       + 2 C_{\alpha\beta} \right) d x^\alpha d x^\beta,
   \label{metric}
\eea
where $a(\eta)$ is a cosmic scale factor. 
$C_{\alpha\beta} ({\bf x},\eta)$ is a transverse and tracefree perturbed 
order tensor variable based on a metric $\delta_{\alpha\beta}$. 
It represents the gravitational wave with $C^{\alpha}_{\alpha} = 0 
= C^{\alpha}_{\beta , \alpha}$, thus having two independent components
which indicate the two polarization states of the gravitational wave.
$C_{\alpha\beta}$ is invariant under the gauge transformation.
The inverse metric, the connection, and the curvatures based on the metric 
in Eq. (\ref{metric}), which are valid to the linear order
in $C_{\alpha\beta}$, are summarized in the Appendix A.
The Appendix A contains all quantities we need for analysing
Eq. (\ref{GFE}) under the metric in Eq. (\ref{metric}) considering to
the linear order in $C_{\alpha\beta}$.
The Appendix B is another presentation using $t$
as the time variable where $dt = a d \eta$. 
$\eta$ and $t$ are the conformal time and the background proper time, 
respectively.
A prime and an overdot indicate the time derivatives based on $\eta$
and $t$, respectively.

{\it 4. Equations for the background and the gravitational wave:}
Using the quantities presented in the Appendix B, Eq. (\ref{GFE}) leads to:
\bea
   & & T^0_0 = - 3 H^2 - 6 \left( B + 3 A \right)
       \left( 2 H \ddot H - \dot H^2 + 6 H^2 \dot H \right),
   \label{AB-eq1} \\
   & & T^0_\alpha = 0 = T^\alpha_0,
   \label{AB-eq2} \\
   & & T^\alpha_\beta = - \delta^\alpha_\beta \Bigg[ 2 \dot H + 3 H^2
   \nonumber \\
   & & \qquad
       + 2 \left( B + 3 A \right) \left( 2 H^{\cdot\cdot\cdot}
       + 12 H \ddot H + 9 \dot H^2 + 18 H^2 \dot H \right) \Bigg]
   \nonumber \\
   & & \qquad
       + \left( 1 + 2 A R \right) D^\alpha_\beta
       + 2 A \dot R \dot C^{\alpha}_{\beta}
   \nonumber \\
   & & \qquad
       - B \Bigg\{ \ddot D^\alpha_\beta + 3 H \dot D^\alpha_\beta
       - \left[ 6 \left( \dot H + H^2 \right) + {\Delta \over a^2} \right]
       D^\alpha_\beta
   \nonumber \\
   & & \qquad
       - 6 \left( \dot H + H^2 \right)^\cdot
       \dot C^{\alpha}_{\beta}
       - 4 \dot H {\Delta \over a^2} C^{\alpha}_{\beta} \Bigg\},
   \label{AB-eq3}
\eea
where $R$ and $D^\alpha_\beta$ are given in Eqs. (\ref{R-def},\ref{D-def}).
The background parts of Eqs. (\ref{AB-eq1},\ref{AB-eq3}) become
($T^a_b \equiv \bar T^a_b + \delta T^a_b$):
\bea
   & & H^2 + 2 \left( B + 3 A \right)
       \left( 2 H \ddot H - \dot H^2 + 6 H^2 \dot H \right) 
       = - {1 \over 3} T^0_0,
   \nonumber \\
   & & \dot H + 2 \left( B + 3 A \right)
       \left( H^{\cdot\cdot\cdot} + 3 H \ddot H + 6 \dot H^2 \right) 
   \nonumber \\
   & & \qquad
       = {1 \over 2} \left( T^0_0 - {1\over 3} \bar T^\alpha_\alpha \right).
   \label{AB-eq-BG}
\eea
The second equation follows from the first one; from $T^a_{b;a} = 0$ we have
$\dot T^0_0 = - H ( 3 T^0_0 - \bar T^\alpha_\alpha )$.
It is remarkable to see that, to the background order, contributions from
$R^{ab} R_{ab}$ term are qualitatively the same as the ones from $R^2$ term.
The perturbed part of Eq. (\ref{AB-eq3}) becomes
\bea
   & & D^\alpha_\beta + 2 A \left( R D^\alpha_\beta 
       + \dot R \dot C^{\alpha}_{\beta} \right)
   \nonumber \\
   & & \quad
       - B \Bigg\{ \ddot D^\alpha_\beta + 3 H \dot D^\alpha_\beta
       - {\Delta \over a^2} \left( D^\alpha_\beta
       + 4 \dot H C^{\alpha}_{\beta} \right) 
   \nonumber \\
   & & \quad 
       - 6 \Bigg[ \left( \dot H + H^2 \right) D^\alpha_\beta
       + \left( \dot H + H^2 \right)^\cdot
       \dot C^{\alpha}_{\beta} \Bigg] \Bigg\} = \delta T^\alpha_\beta,
   \nonumber \\
   & & D^\alpha_\beta \equiv \ddot C^{\alpha}_{\beta}
       + 3 H \dot C^{\alpha}_{\beta} - {\Delta \over a^2} C^{\alpha}_{\beta}.
   \label{AB-eq-GW}
\eea
Equation (\ref{AB-eq-GW}) constitutes a fourth order differential
equation for $C^\alpha_\beta ({\bf x}, t)$ which describes the evolution of 
the gravitational wave.

{\it 5. Case for $R^2$ gravity:}
In the $R^2$ gravity we let $B = 0$.
Using $F \equiv 1 + 2 AR$, Eq. (\ref{AB-eq-BG}) becomes:
\bea
   & & H^2 = - H {\dot F \over F} + {R(F-1) \over 12 F} - {1 \over 3F} T^0_0,
   \nonumber \\
   & & \dot H = - {\ddot F - H \dot F \over 2 F} 
       + {1 \over 2F} \left( T^0_0 - {1\over 3} \bar T^\alpha_\alpha \right).
   \label{A-eq-BG}
\eea
The second equation follows from the first one.
In fact, by re-interpreting $F$ as $F \equiv 1 + 2 ( A + {1 \over 3} B) R$, 
Eq. (\ref{AB-eq-BG}) becomes exactly the same as Eq. (\ref{A-eq-BG}).
Equation (\ref{AB-eq-GW}) becomes
\bea
   \ddot C^{\alpha}_{\beta} + \left( 3 H + {\dot F \over F} \right)
       \dot C^{\alpha}_{\beta} - {\Delta \over a^2} C^{\alpha}_{\beta} 
       = {1 \over F} \delta T^\alpha_\beta.
   \label{A-eq-GW}
\eea
In the large scale limit without the source term
we have a general integral form solution
\bea
   C^\alpha_\beta ({\bf x}, t) = c^\alpha_\beta ({\bf x}) 
       + d^\alpha_\beta ({\bf x}) \int_0^t {dt \over a^3 F},
   \label{LS-sol}
\eea
where the integration constants $c^\alpha_\beta ({\bf x})$ and 
$d^\alpha_\beta ({\bf x})$ indicate the spatial structures in the growing 
mode and decaying mode, respectively.
Apparently, in the large scale limit, Eq. (\ref{AB-eq-GW})
also has a general solution for $C^\alpha_\beta$ which is constant in time.
Pioneering studies of the cosmological gravitational wave
in Eintein gravity are presented in \cite{GW-Einstein}.
Einstein gravity corresponds to a limit of $A = 0 = B$, thus $F = 1$.
The equations for the background and the gravitational wave which are valid 
in a more general context of $f(\phi,R)$ ($\phi$ is a scalar or dilaton field)
gravity were derived in \cite{GGT-1990}; see Eqs. (26,39) in \cite{GGT-1990}
and Eq. (102) in \cite{H-PRW}.
Equation (\ref{A-eq-GW}) remains valid in these types of more generalized
gravity theories; compare with Eq. (26) in \cite{GGT-1990}.
Cosmological studies in these types of generalized gravity theories
can be found in \cite{GGT-others}.

{\it 6. Discussions:}
We have derived the equations describing the evolution of the background 
universe and the gravitational wave in the gravity modified 
by considering the quadratic order correction terms in curvatures.
Such correction terms generically appear as the one-loop quantum 
correction in the investigation of the quantum aspects of the gravity
\cite{QFCS}.
It is known that general $f(R)$ term in the action can be transformed to
the Einstein action by a conformal transformation of the metric.
The basic reason why we obtain a second order differential equation for the 
gravitational wave in the $R^2$ gravity may be traced to the underlying 
symmetry under the conformal transformation, \cite{GGT-1990,GGT-CT}.
However, $R^{ab} R_{ab}$ term cannot be transformed to the Einstein one.

Equation (\ref{AB-eq-BG}) describes the evolution of the background 
universe, and Eq. (\ref{AB-eq-GW}) describes the evolution of 
the gravitational wave under such a background model.
Although the contributions to the background model are qualitatively
the same for $R^2$ and $R^{ab} R_{ab}$ terms, contributions to the
evolution of the gravitational wave are different.
The similar contributions to the background equation are, in fact,
an unexpected and nontrivial result. 
We hope to pay more attention to consequences of 
Eqs. (\ref{AB-eq-BG},\ref{AB-eq-GW}) in future investigations.

Recently the string theories attracted much attention as the potential 
candidates for a successful quantum gravity.
The generic low energy effective action of the string theories
is represented by an action modified by a dilaton field.
The equation for the gravitational wave in such a case is described by an 
equation which is effectively the same as Eq. (\ref{A-eq-GW}), 
see \cite{GGT-1990}.
However, the stringy quantum correction terms are more general
than our action in Eq. (\ref{action}); it yields coupling terms
involving various combinations of the curvatures, the dilaton field,
and their derivatives \cite{string-correction}.
Similar investigations in such a more general gravity theories
will be considered in future.

Results for the scalar and vector type perturbations and applications 
of our results to the relevant cosmological situations will be presented 
in future publications.

We thank Prof. S.-J. Rey for his thoughtful comments.
This work was supported by the KOSEF, Grants No. 95-0702-04-01-3 and 
No. 961-0203-013-1, and through the SRC program of SNU-CTP.

\section*{Appendix A: Linear order quantities}

\noindent
The inverse metric ($0 = \eta$):
\bea
   g^{00} = - {1 \over a^2}, \quad
       g^{0\alpha} = 0, \quad
       g^{\alpha\beta} = {1 \over a^2} \left( \delta^{\alpha\beta}
       - 2 C^{\alpha\beta} \right).
\eea
The connections:
\bea
   & & \Gamma^0_{00} = {a^\prime \over a}, \quad
       \Gamma^0_{0\alpha} = 0, \quad
       \Gamma^0_{\alpha\beta} = {a^\prime \over a} \delta_{\alpha\beta}
       + C^{\prime}_{\alpha\beta} + 2 {a^\prime \over a} C_{\alpha\beta},
   \nonumber \\
   & & \Gamma^\alpha_{00} = 0, \quad
       \Gamma^\alpha_{0\beta} = {a^\prime \over a} \delta^\alpha_\beta
       + C^{\alpha\prime}_{\beta}, 
   \nonumber \\
   & & 
       \Gamma^\alpha_{\beta\gamma} = C^{\alpha}_{\beta,\gamma}
       + C^{\alpha}_{\gamma,\beta} - C^{\;\;\;\; |\alpha}_{\beta\gamma}.
\eea
Riemann curvature:
\bea
   & & R^a_{\;\;b00} = 0 = R^0_{\;\;0ab},
   \nonumber \\
   & & R^0_{\;\;\alpha 0 \beta}
       = \left( {a^\prime \over a} \right)^\prime
       \delta_{\alpha\beta} + C^{\prime\prime}_{\alpha\beta}
       + {a^\prime \over a} C^{\prime}_{\alpha\beta}
       + 2 \left( {a^\prime \over a} \right)^\prime C_{\alpha\beta},
   \nonumber \\
   & & R^0_{\;\;\alpha\beta\gamma}
       = C^{\prime}_{\alpha\gamma,\beta} - C^{\prime}_{\alpha\beta,\gamma},
   \nonumber \\
   & & R^\alpha_{\;\; 00\beta}
       = \left( {a^\prime \over a} \right)^\prime
       \delta^\alpha_\beta + C^{\alpha\prime\prime}_{\beta}
       + {a^\prime \over a} C^{\alpha\prime}_{\beta},
   \nonumber \\
   & & R^\alpha_{\;\; 0 \beta\gamma}
       = C^{\alpha\prime}_{\gamma,\beta}
       - C^{\alpha\prime}_{\beta,\gamma}, \quad
       R^\alpha_{\;\; \beta 0\gamma}
       = C^{\alpha\prime}_{\gamma,\beta}
       - C^{\prime\;\;\; |\alpha}_{\beta\gamma},
   \nonumber \\
   & & R^\alpha_{\;\;\beta\gamma\delta}
       = \left( {a^\prime \over a} \right)^2 \left( \delta_{\beta\delta}
       \delta^\alpha_\gamma - \delta_{\beta\gamma} 
       \delta^\alpha_\delta \right)
   \nonumber \\
   & & \qquad
       + {a^\prime \over a} \left(
       \delta_{\beta\delta} C^{\alpha\prime}_{\gamma}
       - \delta_{\beta\gamma} C^{\alpha\prime}_{\delta}
       + \delta^\alpha_\gamma C^{\prime}_{\beta\delta}
       - \delta^\alpha_\delta C^{\prime}_{\beta\gamma} \right)
   \nonumber \\
   & & \qquad
       + 2 \left( {a^\prime \over a} \right)^2
       \left( \delta^\alpha_\gamma C_{\beta\delta}
       - \delta^\alpha_\delta C_{\beta\gamma} \right)
   \nonumber \\
   & & \qquad
       + C^{\alpha}_{\delta,\beta\gamma} - C^{\alpha}_{\gamma,\beta\delta}
       - C^{\;\;\;\; |\alpha}_{\beta\delta \;\;\;\; \gamma}
       + C^{\;\;\;\; |\alpha}_{\beta\gamma \;\;\;\; \delta}.
\eea
Ricci curvature:
\bea
   & & R^0_0 = {3 \over a^2} \left( {a^\prime \over a} \right)^\prime, \quad
       R^0_\alpha = 0 = R^\alpha_0,
   \nonumber \\
   & & R^\alpha_\beta = {1 \over a^2} \Bigg\{
       \left[ \left( {a^\prime \over a} \right)^\prime
       + 2 \left( {a^\prime \over a} \right)^2 \right] \delta^\alpha_\beta
   \nonumber \\
   & & \qquad
       + C^{\alpha\prime\prime}_{\beta}
       + 2 {a^\prime \over a} C^{\alpha\prime}_{\beta}
       - \Delta C^{\alpha}_{\beta} \Bigg\}.
\eea
Scalar curvature:
\bea
   R = {6 \over a^2} \left[ \left( {a^\prime \over a} \right)^\prime
       + \left( {a^\prime \over a} \right)^2 \right].
\eea
Einstein tensor:
\bea
   & & G^0_0 = - {3 \over a^2} \left( {a^\prime \over a} \right)^2, \quad
       G^0_\alpha = 0 = G^\alpha_0,
   \nonumber \\
   & & G^\alpha_\beta = {1 \over a^2} \Bigg\{
       - \left[ 2 \left( {a^\prime \over a} \right)^\prime
       + \left( {a^\prime \over a} \right)^2 \right] \delta^\alpha_\beta
   \nonumber \\
   & & \qquad
       + C^{\alpha\prime\prime}_{\beta}
       + 2 {a^\prime \over a} C^{\alpha\prime}_{\beta}
       - \Delta C^{\alpha}_{\beta} \Bigg\}.
\eea
Curvature combinations:
\bea
   & & R^a_b R^b_a = {12 \over a^4} \left[
       \left( {a^\prime \over a} \right)^{\prime 2}
       + \left( {a^\prime \over a} \right)^\prime
       \left( {a^\prime \over a} \right)^2
       + \left( {a^\prime \over a} \right)^4 \right],
   \\
   & & R^{cd} R^0_{\;\; cd0} = - {3 \over a^4} \left( {a^\prime \over a}
       \right)^\prime \left[ \left( {a^\prime \over a} \right)^\prime
       + 2 \left( {a^\prime \over a} \right)^2 \right],
   \nonumber \\
   & & R^{cd} R^0_{\;\; cd\alpha} = 0 = R^{cd} R^\alpha_{\;\; cd0},
   \nonumber \\
   & & R^{cd} R^\alpha_{\;\; cd\beta} = {1 \over a^4} \Bigg\{ 
   \nonumber \\
   & & \qquad
       - \left[ 3 \left( {a^\prime \over a}
       \right)^{\prime 2} + 2 \left( {a^\prime \over a} \right)^\prime
       \left( {a^\prime \over a} \right)^2
       + 4 \left( {a^\prime \over a} \right)^4 \right] \delta^\alpha_\beta
   \nonumber \\
   & & \qquad
       + \left[ - 3 \left( {a^\prime \over a} \right)^\prime
       + \left( {a^\prime \over a} \right)^2 \right]
       C^{\alpha\prime\prime}_{\beta}
       - 4 {a^\prime \over a} \left( {a^\prime \over a} \right)^\prime
       C^{\alpha\prime}_{\beta}
   \nonumber \\
   & & \qquad
       + \left[ \left( {a^\prime \over a} \right)^\prime
       + \left( {a^\prime \over a} \right)^2 \right]
       \Delta C^{\alpha}_{\beta} \Bigg\}.
\eea
Covariant derivatives:
\bea
   & & R^{;0}_{\;\;\; 0} = - {1 \over a^2} \left(
       R^{\prime\prime} - {a^\prime \over a} R^\prime \right), \quad
       R^{;0}_{\;\;\;\alpha} = 0 = R^{;\alpha}_{\;\;\; 0}, 
   \nonumber \\
   & & 
       R^{;\alpha}_{\;\;\;\beta} = - {1 \over a^2}
       \left( {a^\prime \over a} \delta^\alpha_\beta
       + C^{\alpha\prime}_{\beta} \right) R^\prime,
   \\
   & & \Box R = - {1 \over a^2} \left( R^{\prime\prime}
       + 2 {a^\prime \over a} R^\prime \right),
   \\
   & & \Box R^0_0 = - {1 \over a^2} \left( R^{0\prime\prime}_0
       + 2 {a^\prime \over a} R^{0\prime}_0 \right) 
   \nonumber \\
   & & \qquad
       + 12 {1 \over a^4} \left( {a^\prime \over a} \right)^2
       \left[ \left( {a^\prime \over a} \right)^\prime
       - \left( {a^\prime \over a} \right)^2 \right],
   \nonumber \\ 
   & & \Box R^0_\alpha = 0 = \Box R^\alpha_0,
   \nonumber \\
   & & \Box R^\alpha_\beta
       = - {1 \over a^2} \left[ \bar R^{\alpha \prime\prime}_\beta
       + 2 {a^\prime \over a} \bar R^{\alpha \prime}_\beta \right)
   \nonumber \\
   & & \qquad 
       - 4 \delta^\alpha_\beta {1 \over a^4} 
       \left( {a^\prime \over a} \right)^2
       \left[ \left( {a^\prime \over a} \right)^\prime
       - \left( {a^\prime \over a} \right)^2 \right],
   \nonumber \\
   & & \qquad
       - {1 \over a^2} \Bigg[ D^{\alpha\prime\prime}_\beta
       + 2 {a^\prime \over a} D^{\alpha\prime}_\beta
       - 2 \left( {a^\prime \over a} \right)^2 D^\alpha_\beta
       - \Delta D^\alpha_\beta \Bigg]
   \nonumber \\
   & & \qquad
       - 8 {1 \over a^4} {a^\prime \over a} \left[
       \left( {a^\prime \over a} \right)^\prime
       - \left( {a^\prime \over a} \right)^2 \right] C^{\alpha\prime}_{\beta},
\eea
where
\bea
   D^\alpha_\beta \equiv
       {1 \over a^2} \left( C^{\alpha\prime\prime}_{\beta}
       + 2 {a^\prime \over a} C^{\alpha\prime}_{\beta}
       - \Delta C^{\alpha}_{\beta} \right),
\eea
and an overbar indicates a background order of the variable.

\vskip .5cm
\centerline{\bf APPENDIX B: IN TERMS OF $t$}

\vskip .5cm
Using $t$ as the time variable we have ($H \equiv \dot a/a$):
\bea
   & & R^0_0 = 3 \left( \dot H + H^2 \right), \quad
       R^0_\alpha = 0 = R^\alpha_0,
   \nonumber \\
   & & R^\alpha_\beta = \left( \dot H + 3 H^2 \right) \delta^\alpha_\beta
       + \ddot C^{\alpha}_{\beta} + 3 H \dot C^{\alpha}_{\beta}
       - {\Delta \over a^2} C^{\alpha}_{\beta},
   \\
   & & R = 6 \left( \dot H + 2 H^2 \right),
   \label{R-def}
   \\
   & & G^0_0 = - 3 H^2, \quad G^0_\alpha = 0 = G^\alpha_0,
   \nonumber \\
   & & G^\alpha_\beta = - \left( 2 \dot H + 3 H^2 \right) \delta^\alpha_\beta
       + \ddot C^{\alpha}_{\beta} + 3 H \dot C^{\alpha}_{\beta}
       - {\Delta \over a^2} C^{\alpha}_{\beta},
   \\
   & & R^a_b R^b_a = 12 \left( \dot H^2 + 3 \dot H H^2 + 3 H^4 \right),
   \\
   & & R^{cd} R^0_{\;\; cd0} = - 3 \left( \dot H + H^2 \right)
       \left( \dot H + 3 H^2 \right), 
   \nonumber \\
   & & 
       R^{cd} R^0_{\;\; cd\alpha} = 0 = R^{cd} R^\alpha_{\;\; cd0},
   \nonumber \\
   & & R^{cd} R^\alpha_{\;\;cd\beta}
       = - \left( 3 \dot H^2 + 8 \dot H H^2 + 9 H^4 \right) 
       \delta^\alpha_\beta
   \nonumber \\
   & & \qquad
       - \left( 3 \dot H + 2 H^2 \right) \ddot C^{\alpha}_{\beta}
       - \left( 7 \dot H + 6 H^2 \right) H \dot C^{\alpha}_{\beta}
   \nonumber \\
   & & \qquad
       + \left( \dot H + 2 H^2 \right) {\Delta \over a^2} C^{\alpha}_{\beta},
   \\
   & & R^{;0}_{\;\;\; 0} = - \ddot R, \quad
       R^{;0}_{\;\;\;\alpha} = 0 = R^{;\alpha}_{\;\;\; 0}, \quad
   \nonumber \\
   & & 
       R^{;\alpha}_{\;\;\;\beta} = - \left( H \delta^\alpha_\beta
       + \dot C^{\alpha}_{\beta} \right) \dot R,
   \\
   & & \Box R = - \left( \ddot R + 3 H \dot R \right),
   \\
   & & \Box R^0_0 = - 3 \left( H^{\cdot\cdot\cdot} + 5 H \ddot H + 2 \dot H^2
       + 2 H^2 \dot H \right), 
   \nonumber \\
   & &
       \Box R^0_\alpha = 0 = \Box R^\alpha_0,
   \nonumber \\
   & & \Box R^\alpha_\beta = - \delta^\alpha_\beta \left(
       H^{\cdot\cdot\cdot} + 9 H \ddot H + 6 \dot H^2 + 22 H^2 \dot H \right)
   \nonumber \\
   & & \qquad
       - \left( \ddot D^\alpha_\beta + 3 H \dot D^\alpha_\beta
       - 2 H^2 D^\alpha_\beta - {\Delta \over a^2} D^\alpha_\beta \right)
   \nonumber \\
   & & \qquad
       - 8 H \dot H \dot C^{\alpha}_{\beta},
   \\
   & & D^\alpha_\beta \equiv \ddot C^{\alpha}_{\beta}
       + 3 H \dot C^{\alpha}_{\beta} - {\Delta \over a^2} C^{\alpha}_{\beta}.
   \label{D-def}
\eea


\end{document}